# Partially temporal coherent spatiotemporal Laguerre-Gaussian and its propagation dynamics

Mengxiang Wu[1,†], Youngbin Park[2,3,4†], Nianjia Zhang[1], Taemin Son[2,3,4], Qiwen Zhan[5,6], Andy Chong[2,3,4,*], and Qian Cao[1,*]

[1] *School of Optical-Electrical and Computer Engineering, University of Shanghai for Science and Technology, Shanghai 200093, China*

[2] *Department of Physics, Pusan National University, Busan, 46241, Republic of Korea*

[3] *Institute for Future Earth, Pusan National University, Busan 46241, Republic of Korea*

[4] *Quantum Science Technology Center, Pusan National University, Busan 46241, Republic of Korea*

[5] *Zhejiang Key Laboratory of 3D Micro/Nano Fabrication and Characterization, Department of Electronic and Information Engineering, School of Engineering, Westlake University, Hangzhou 310030, China*

[6] *Westlake-Kromoson Joint Laboratory of Light-Field Intelligence, School of Engineering, Westlake University, Hangzhou 310030, China*

* *cao.qian@usst.edu.cn, chong0422@pusan.ac.kr*

## Abstract

Optical spatiotemporal vortices carrying transverse photonic orbital angular momentum have become an intense research topic in the recent study of structured light fields. Despite being mainly studied within the framework of fully coherent optical pulses, the complex topologies of the light field are also investigated with partially coherent light sources, offering an alternative pathway in applications. In this work, we theoretically and experimentally investigate partially temporally coherent spatiotemporal Laguerre-Gaussian (STLG) field with independently controllable radial and azimuthal indices. The spatiotemporal topological information, carried by the structured amplified spontaneous emission source, are well preserved in the presence of inherent instantaneous random temporal phase fluctuations. Under combined spatiotemporal astigmatism and group velocity dispersion, the partially temporally coherent STLG field can be converted into spatiotemporal Hermite-Gaussian (STHG) field. Both structures maintain their spatiotemporal characteristics after propagating through scattering media. Partially temporally coherent structured light fields may offer new insights into photonic topology and complex wave dynamics, with potential applications in high-fidelity optical communications and high-dimensional information processing in highly scattering environments.

## Introduction

For decades, light beams carrying orbital angular momentum (OAM) have attracted intense research interest, mainly in the study of optical vortex beams possessing helical wavefronts with spatial phase singularities. Since the seminal discovery by Allen et al. [1] in 1992 that light beams with a spiral phase factor of $\exp(il\phi)$ carry a longitudinal OAM of $l\hbar$ per photon, extensive progress has been made in unveiling the physical properties of such optical vortices. These spatial vortex beams are typically described by Laguerre-Gaussian (LG) modes with azimuthal ($l$) and radial ($p$) indices and have proven to be of significant value in a wide range of applications, including high-capacity optical communications [2], super-resolution microscopy [3,4], optical tweezing [5], and high-dimensional quantum information processing [6].

In 2020, a novel class of optical vortices featuring transverse OAM, known as spatiotemporal optical vortices (STOV) [7-10], has emerged and been successfully generated. Distinct from conventional optical vortex beams [11], STOV wavepackets exhibit spiral phases and phase singularities within the spatiotemporal domain. Due to their unique spatiotemporal coupling characteristics, STOVs hold potential in applications such as spatiotemporal differentiation and dynamic light-matter interactions. The study on STOV wavepacket mainly focuses on fully coherent optical pulses emitted from mode-locked lasers. In many practical applications, partially coherent light sources can be advantageous in many applications, for example, in enhancing the signal-to-noise ratio and mitigating speckle noise [12]. Recently, fundamental STOVs with partial temporal coherence have been investigated, while the study has been confined to simple vortex states [13]. Despite high-order spatiotemporal Laguerre-Gaussian (STLG) and Hermite-Gaussian (STHG) were studied in the context of fully coherent laser sources[14], none report has been made in generating and studying those complicated ST fields from partially temporal coherent light fields.

Recent advances have also explored fractional STOVs [15], 4Pi-generated STOV arrays [16], spatiotemporal coherence engineering [17], and three-dimensional autofocusing with Airy pulses [18], highlighting the growing diversity of spatiotemporal structured light. Here, we present a theoretical and experimental investigation of the partially temporally coherent STLG light field. We theoretically elucidate the formation mechanism of these high-order topological structures in the presence of random phase fluctuations. Experimentally, we demonstrate the generation of partially coherent STLG with independently controllable radial and azimuthal indices. Its conversion into partially temporal coherent STHG light field is demonstrated under the combined influence of spatiotemporal astigmatism and group velocity dispersion. Besides, these light fields also exhibit robustness during propagation inside a scattering medium. Our findings reveal that, despite the inherent instantaneous intensity fluctuations of the spontaneous emission light source, partially coherent ST structured light fields can still preserve complex spatiotemporal correlations, thus offering a promising platform for robust data transmission and potential particle manipulation, especially when combined with low-cost and readily available optical sources.

## Theoretical analysis and simulations

To describe the formation and propagation of partially temporally coherent spatiotemporal Laguerre-Gaussian (STLG) light field, we first define the field in its spatial frequency and optical frequency $(k_x,\ \omega)$ domain. Here, $\Omega = \omega - \omega_0$ represents the reduced angular frequency (with $\omega_0$ being the central frequency) and $k_x$ denotes the transverse spatial frequency. In this domain, we define the dimensionless ST polar coordinates $(r, \theta)$ as: $r = \sqrt{k_x^2 + \gamma^2 \Omega^2}$, $\theta = \tan^{-1}(\gamma\Omega/k_x)$, where $\gamma$ is a scaling factor introduced to normalize the dimensions of the spatial and optical frequency components. Its inverse, $\alpha = \gamma^{-1}$, will later appear in the ST domain definition. Based on the concept of partial temporal coherence, the STLG light field generated from a partially temporal coherent source, for example, amplified spontaneous emission (ASE), can be numerically simulated by applying a randomly distributed spectral phase term, $e^{i\phi_{\mathrm{rand}}(\Omega)}$. Consequently, the target frequency-domain complex-amplitude distribution $\tilde{E}(k_x, \Omega)$ is modeled as:

$$\tilde{E}(k_x, \Omega) = \left[g_{p,l}(r) e^{-il\theta}\right] \cdot e^{i\phi_{\mathrm{rand}}(\Omega)}, \tag{1}$$

where $\phi_{\mathrm{rand}}(\Omega) \sim N(0, \sigma^2)$ follows a normal distribution with variance $\sigma^2$. When $\sigma^2 = 0$, it represents a perfectly mode-locked pulse, while an increasing $\sigma^2$ characterizes the transition to a partially temporal coherent random pulse. The radial basis function $g_{p,l}(r)$ is given by:

$$g_{p,l}(r) = C_{p,l} \left(\frac{r}{w_1}\right)^{|l|} L_p^{|l|}\left(\frac{2r^2}{w_1^2}\right) \exp\left(-\frac{r^2}{w_1^2}\right), \tag{2}$$

where $C_{p,l}$ is the normalization constant and $w_1$ is the spectral waist radius.

The field distribution $E(x,t)$ in the ST domain can be derived via a two-dimensional (2D) spatiotemporal Fourier transform of Eq. (1). Because the random phase $\phi_{\mathrm{rand}}(\Omega)$ depends only on $\Omega$ (and not on $k_x$), the 2D Fourier transform can be viewed as the convolution between a Hankel transform (for the vortex structure) and a temporal Fourier transform (for the stochastic envelope). The field in ST polar coordinates $(\rho,\phi)$ is thus expressed as:

$$E(\rho,\phi) \approx \left[2\pi(-i)^l e^{-il\phi}\mathcal{H}_l\{g_{p,l}(\rho)\}\right] * \xi(t) \tag{3}$$

where the symbol "*" denotes the convolution operator. The conjugated ST coordinates $(\rho,\phi)$ are defined as $\rho = \sqrt{x^2+\alpha^2 t^2}$ , $\phi = \tan^{-1}(\alpha t/x)$ with $\alpha = \gamma^{-1}$ . Here, $\mathcal{H}_l\{g_{p,l}(\rho)\} = \int_0^\infty r\, g_{p,l}(r) J_l(2\pi\rho r)dr$ represents the $l$-th order Hankel transform of the radial function, $J_l$ is the Bessel function of the first kind. $\xi(t) = FT\{e^{i\phi_{\mathrm{rand}(\Omega)}}\}$ is the temporal random modulation function induced by the random spectral phase, representing the temporal phase fluctuations of the partially coherent light field. This mathematical description shows that the random spectral phase enters only through the stochastic temporal modulation $\xi(t)$, while the deterministic spatial vortex phase factor $e^{-il\phi}$ remains explicitly present in the field expression. The full ST structure is therefore not directly destroyed by the random phase, although its observable robustness must be further examined through the correlation function.

To test the robustness of the ST topological structure against the randomness of the temporal phase, we simulated the STLG mode with radial index $p = 1$ and azimuthal index $l = +1$ under varying degrees of partial temporal coherence ($\sigma^2 = 0, \pi^2/16$, and $\pi^2$). The $x = 0$ temporal intensity profiles (Figs. 1(a1–a3)) show that the introduction of spectral phase noise induces rapid amplitude fluctuations and random modulations in the temporal domain. More revealingly, the 3D iso-intensity surfaces of the light field (Figs. 1 (b1–b3)) and the corresponding 2D ST cross-sections (Figs. 1(c1–c3)) illustrate the structural evolution of the ST light field with varying phase randomness. Specifically, when $\sigma^2 = 0$, the field corresponds to a fully temporally coherent source, exhibiting a pristine STLG profile with smooth concentric rings and a well-defined central singularity. As the phase variance increases to $\sigma^2 = \pi^2/16$, representing a partially temporally coherent state, the spatiotemporal envelope experiences noticeable localized distortions and intensity modulations, yet the macroscopic ring geometry remains identifiable. Finally, when $\sigma^2 = \pi^2$, simulating a low-coherence source, the instantaneous spatiotemporal envelope is severely disrupted and highly fragmented into random speckle-like patterns. However, our later experimental findings would confirm the presence of complicated ST correlation structures, such as STLG and STHG, under the strong temporal phase fluctuations from the ASE light source. The partially temporally coherent ST structured light field can also retain its topological information, making such fields promising for applications involving ASE-type sources or turbulent environments where full temporal coherence is unavailable.

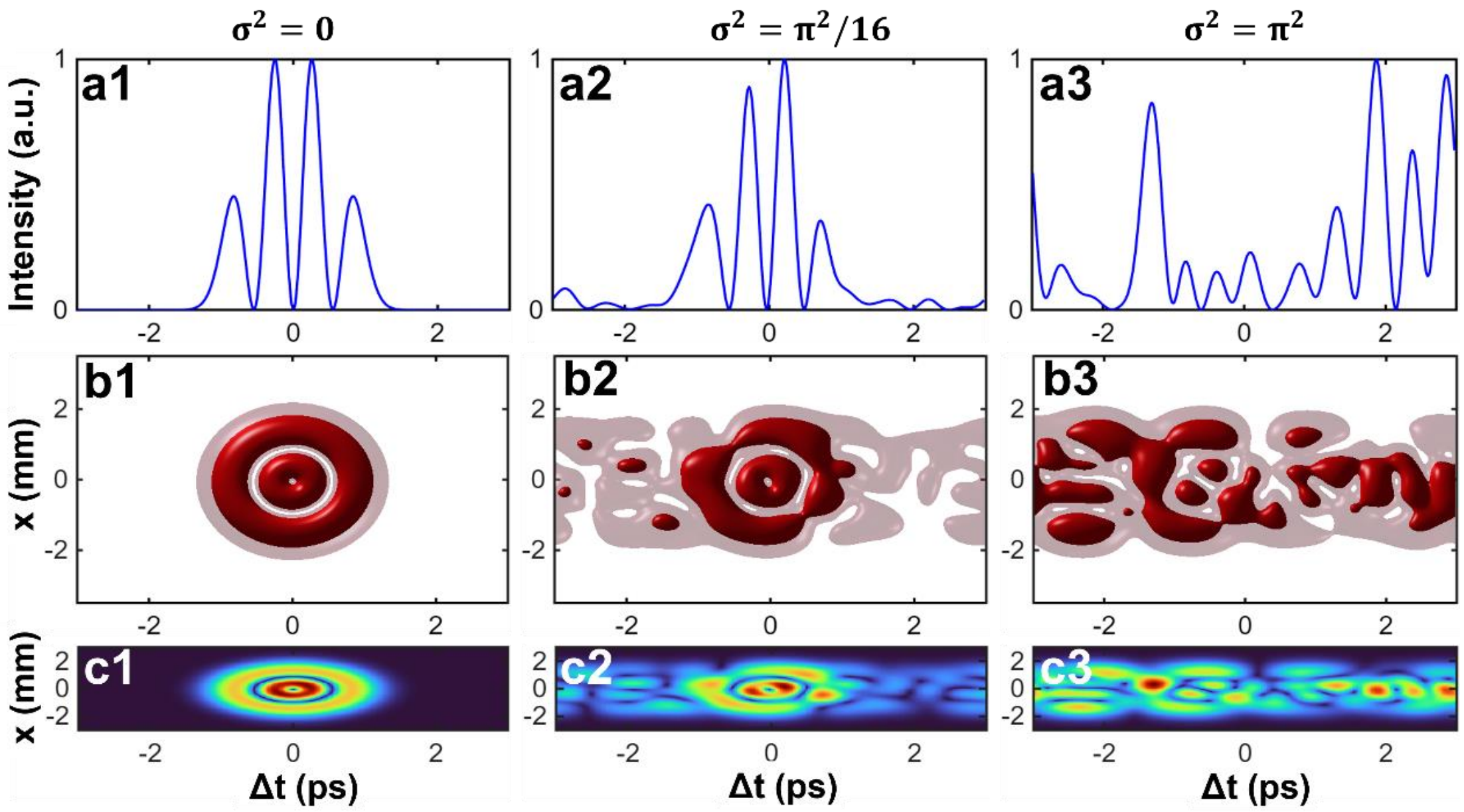

**Figure 1. Numerically simulated profiles of a partially temporally coherent STLG field under different spectral phase noise variances.** The STLG has $p = 1$ and $l = 1$. The central wavelength is 1030 nm with a 3-dB bandwidth of 8 nm. Each column corresponds to different phase variance, showing the one-dimensional temporal intensity profiles (a1 to a3), three-dimensional iso-intensity surfaces (b1 to b3), and two-dimensional ST distributions (c1 to c3). Dark and translucent red surfaces represent 0.1 and 0.01 of the maximum intensity, respectively.

**Experimental results**

We experimentally synthesize the partially temporally coherent STLG light field using a complex amplitude modulation technique like the one in [19]. As illustrated in Figure 2(a), this method integrates a two-dimensional ST light modulation with pulse shaping principles. The spectrum of the incident light is first dispersed by a diffraction grating and then collimated by a cylindrical lens onto a spatial light modulator (SLM). This optical arrangement maps the frequency components across the transverse spatial axis to establish a two-dimensional spatial and spectral domain. Within this plane, we apply a specifically designed computer-generated hologram (CGH) to sculpt the light field in the ST domain. The CGH is generated by combining the phase and amplitude distributions of the target spatial-spectral Laguerre-Gaussian mode with a blazed grating phase. By carefully tuning the modulation depth of this added blazed grating, we can precisely control the amplitude of the ST light.

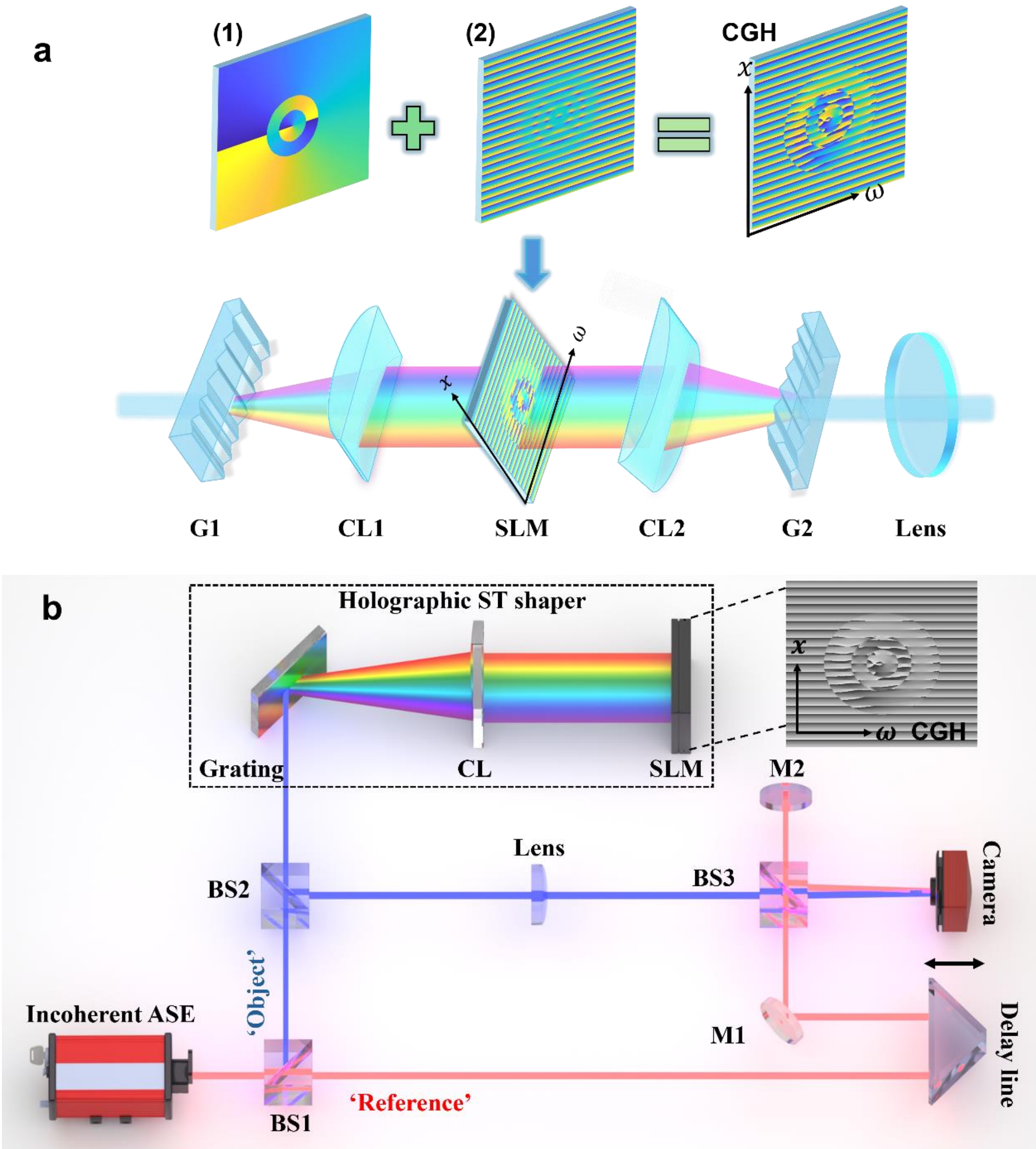

**Figure 2. Experimental schematic and setup for generating and characterizing partially temporally coherent spatiotemporal light fields**. **(a)** Concept of the partially temporally coherent ST light synthesis based on complex-amplitude modulation technique. **(b)** Schematic of the setup for partially temporally coherent structured ST light field generation.

BS, beam splitter; M, mirror; CL, cylindrical lens; SLM, spatial light modulator.

The detailed experimental configuration for generating and characterizing the partially temporally coherent field is presented in Figure 2(b). The light source is a home-built ytterbium-doped fiber ring cavity operating below the lasing threshold, which provides ASE output centered at 1030 nm with a spectral bandwidth of about 4 nm. The initial beam is divided into an object path and a reference path. In the object path, the beam enters a folded two-dimensional holographic ST shaper consisting of a reflective diffraction grating (1200 lines/mm), a cylindrical lens (10cm focal length on the y-axis), and a high-resolution phase-only SLM (Holoeye GAEA-2, 3840×2160 pixels with a pitch of 3.74 μm). The SLM is located in the spatial-spectral Fourier plane and encodes the complex field amplitude of the target

ST wavepacket. Specifically, a one-dimensional blazed phase grating with spatially varying modulation depth is superimposed on the hologram. Unwanted light is diffracted into the first order and angularly discarded, while the desired zeroth-order field is retroreflected along the optical axis and recombined by the grating to synthesize the partially temporally coherent ST structured light. For full characterization, the generated object beam is combined with the unmodulated reference beam at a small tilt angle on a CCD camera, forming off-axis interference fringes. By changing their relative time delay, the three-dimensional information of the generated wavepacket is recorded and reconstructed.

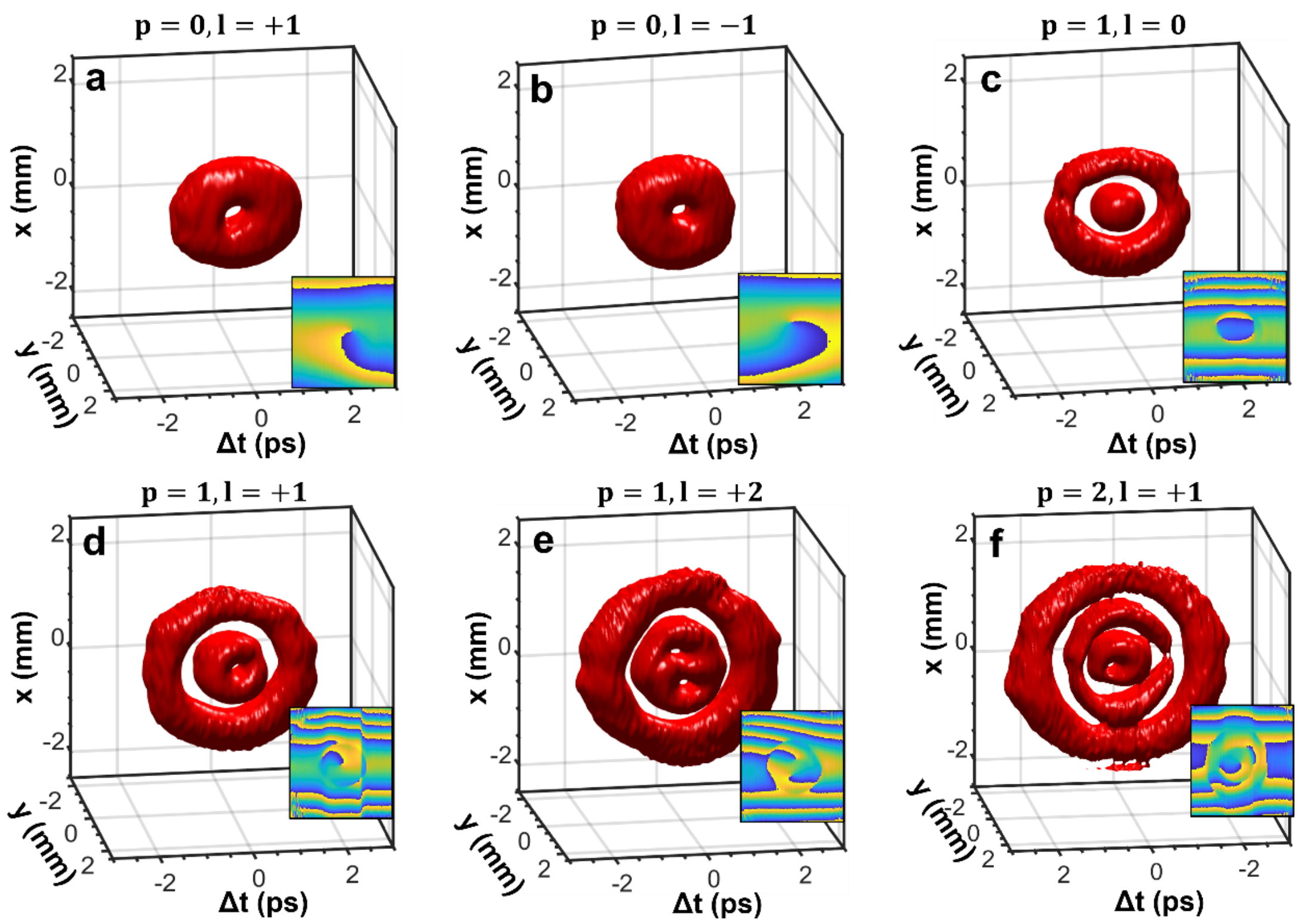

**Figure 3. Experimentally reconstructed partially temporally coherent STLG wavepackets.** The radial and azimuthal indices $(p, l)$ correspond to (a) $(0, +1)$; (b) $(0, -1)$; (c) $(1,0)$; (d) $(1, +1)$; (e) $(1, +2)$; and (f) $(2, +1)$; The iso-value for all surfaces is set to 0.08. The inset panels show the correlated spatiotemporal phase located at the $y = 0$ plane.

To demonstrate the capability of our modulation and measurement schematic, we first generate partially temporal coherent STLG light with different indices. Figure 3 shows the measurement results. The isosurface shown in these plots corresponds to the ST correlation pattern of the modulated STLG light. It can be seen that the ring and multi-ring structure is generated in these partially temporal coherent structured light fields. The correlated ST phase, as shown in the inset figures, confirms the generation of ST vortex phase structure and the STLG phase structure generated from the ASE light source.

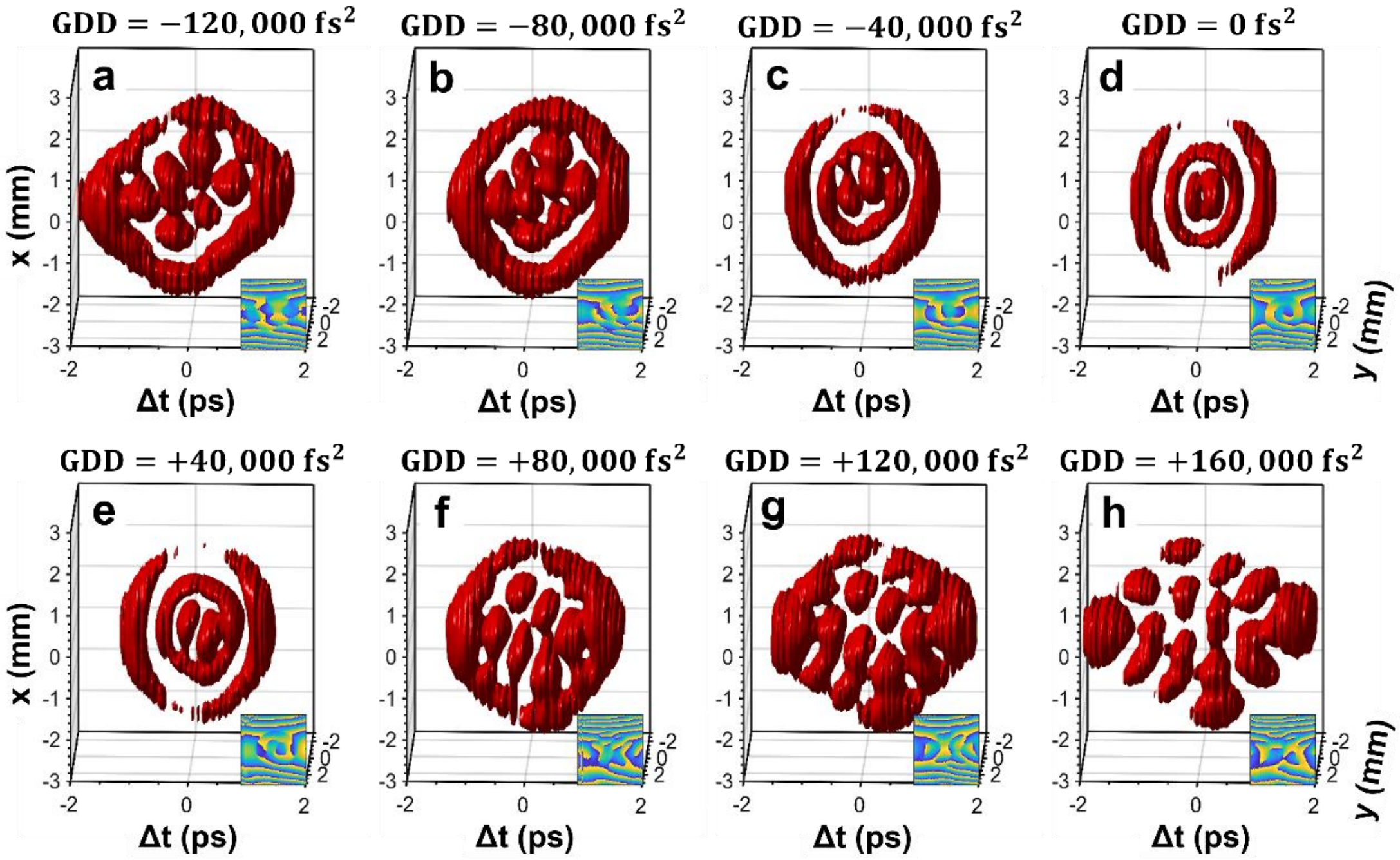

**Figure 4. Experimentally reconstructed correlation structure of the partially temporally coherent ST-HLG wavepackets undergoing different GDD values.** STLG has $p = 2$ and $l = +1$. GDD is set with an interval of 40,000 $fs^2$.The iso-intensity level is set to 0.05 for all cases. Insets show the corresponding 2D ST phase correlation at the $y = 0$ plane.

Furthermore, we investigated the ST astigmatic conversion of high-dimensional partially temporally coherent light fields from STLG to STHG [14]. Figure 4 shows the experimentally reconstructed evolution of a high-order STLG mode with $p = 2$ and $l = +1$ under different group-delay dispersion (GDD) values. At large negative GDD values (figs. 4a and 4b), the original concentric-ring STLG structure gradually deforms, giving rise to an intermediate ST Hermite-Laguerre-Gaussin (STHLG) configuration [20]. As the GDD approaches zero, the concentric-ring topology becomes more symmetric and well defined, reaching its most symmetric form at 0 $fs^2$ (figs. 4d). As the positive GDD increases further (figs. 4e-4h), the ring structure progressively deforms into multiple separated lobes, indicating a continuous evolution through intermediate ST-HLG states toward a high-order STHG-like profile. The observed evolution originates from the quadratic spectral phase introduced by GDD, which induces a dispersive temporal phase modulation of the wavepacket. In combination with ST astigmatism, this dispersive phase modulation continuously reshapes the ST correlation structure, enabling the transformation from the ring-like STLG topology through intermediate ST-HLG states to the multi-lobed STHG structure. These results demonstrate that the ST astigmatic conversion remains robust under partial temporal coherence, with the underlying mode-conversion process preserved despite the presence of temporal phase randomness.

To further investigate the robustness of these partially temporally coherent high-dimensional ST wavepackets in complex, scattering optical environments, we introduced ground glass scattering plates into the object beam path. By utilizing these plates with varying grit numbers (1500 and 600 grit), we effectively simulated different degrees of random spatial phase perturbations applied to these partially temporally coherent light fields. Note it that for these ground glass plates, a lower grit number indicates a coarser surface and should apply a stronger scattering effect.

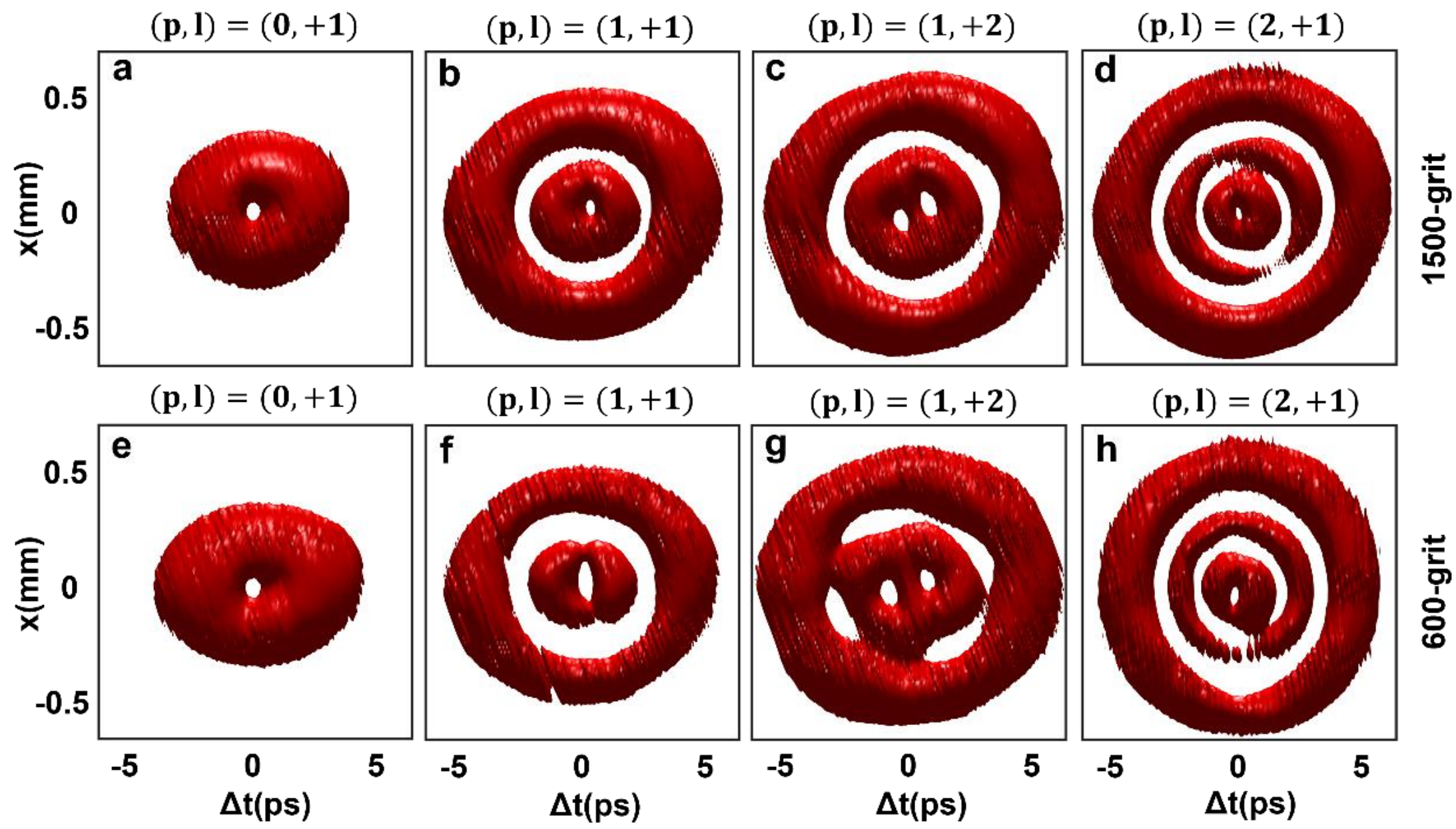

**Figure 5. Experimentally reconstructed correlation structure of partially temporally coherent STLG field with spatial scattering.** The top row (**a** to **d**) displays distinct modes $(p, l) = (0, +1), (1, +1), (1, +2)$, and $(2, +1)$) after passing through a 1500-grit ground glass scattering plate. The bottom row (**e** to **h**) depicts the corresponding modes under stronger scattering conditions utilizing a 600-grit plate.

The first set of results focuses on the partially temporally coherent STLG fields, as illustrated in Figures 5. The experimental reconstructions correspond to the three-dimensional correlation of the light field with four distinct modes corresponding to $(p, l) = (0, +1)$, $(1, +1)$, $(1, +2)$, and $(2, +1)$. Figures 5(a-d) show the correlation pattern of the field after passing through a weaker scattering medium using a 1500-grit plate, while Figures 5(e-h) depict the same modes under stronger scattering conditions using a 600-grit plate. Although the increased random scattering at 600-grit induces a noticeable speckle effect causing the correlation of the spatiotemporal light field become much rougher, the topological features within the ST partially temporal coherent light field are still well preserved, as can be seen in its central vortex core and the multi-ring structure.

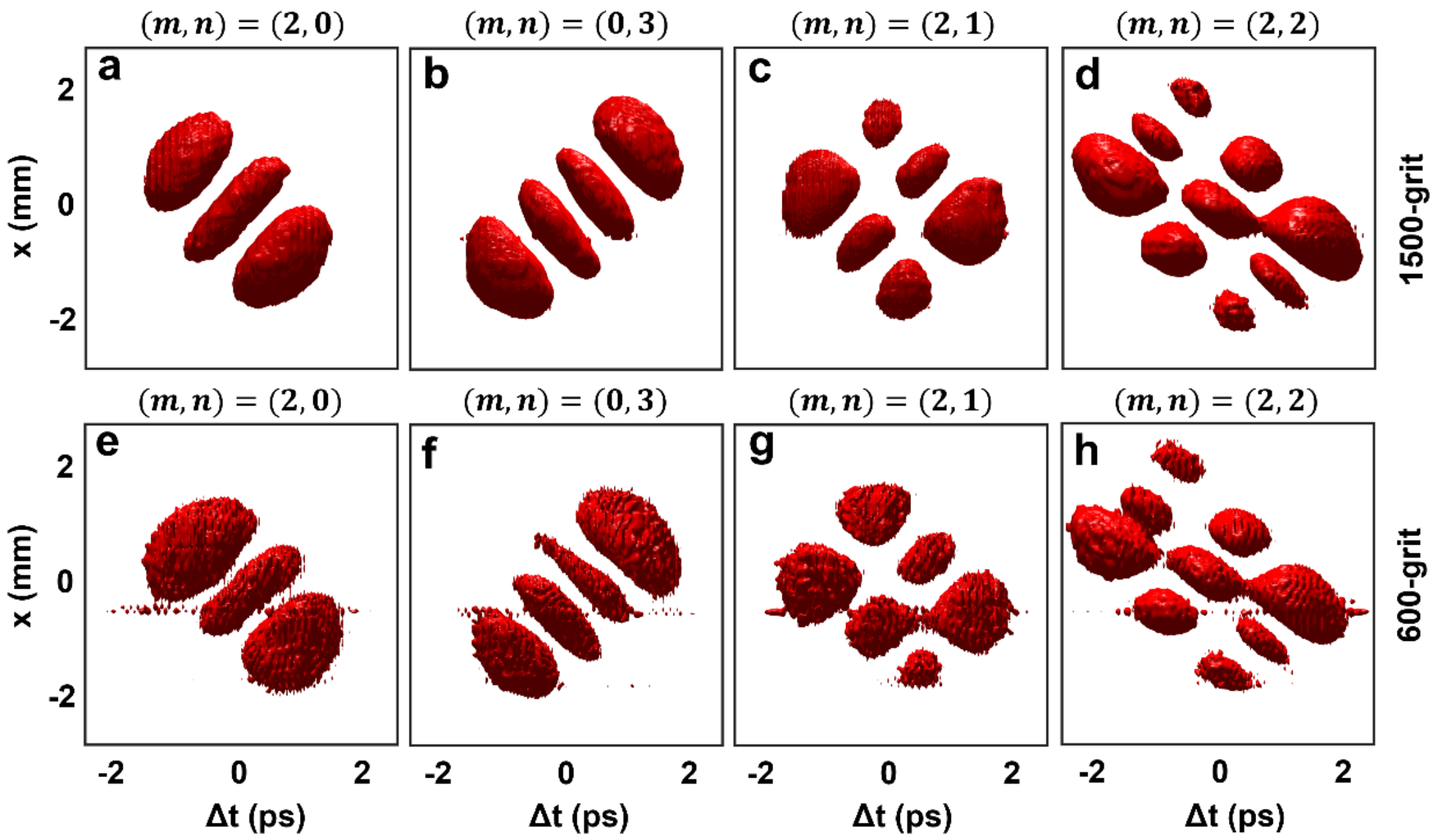

**Figure 6. Experimentally reconstructed 3D iso-intensity surfaces of partially temporally coherent STHG wavepackets under random spatial phase perturbations.** The top row (a-d) displays distinct modes ($(m, n) = (2,0)$, $(0,3)$, $(2,1)$, and $(2,2)$) after passing through a 1500-grit ground glass scattering plate. The bottom row (e-h) depicts the corresponding modes under stronger scattering conditions utilizing a 600-grit plate.

Similarly, we examined the resilience of partially temporally coherent STHG light fields under identical scattering conditions, as presented in Fig. 6. Four different STHG modes corresponding to $(m, n) = (2, 0)$, $(0, 3)$, $(2, 1)$, and $(2, 2)$ were investigated under two scattering settings. Similar to the previous STLG results, as the scattering severity increases from a 1500-grit to a 600-grit diffuser, the isosurface plots show that the shape of the ST correlation structure of the light field become progressively superimposed with fine speckles, resulting in a rougher appearance. Nevertheless, despite this localized speckle noise, the reconstructed three-dimensional correlation wavepackets still clearly retain their characteristic STHG structural signatures. These observations indicate that the spatiotemporal correlation structures of both partially temporally coherent STLG and STHG light fields remain highly resilient after propagation through random scattering media. This inherent robustness significantly enhances their potential for reliable optical communication and robust imaging applications in turbulent or highly scattering environments.

## Conclusion

In conclusion, we have successfully generated partially temporally coherent spatiotemporal Laguerre-Gaussian (STLG) light field with independently controllable radial and azimuthal indices. Furthermore, we demonstrated the spatiotemporal astigmatic conversion of STLG modes into spatiotemporal Hermite-Gaussian (STHG) modes. Both simulation and experiment results confirms the generation of complex ST light field from partially temporal coherent light sources. We also verified the exceptional resilience of these high-dimensional macroscopic topologies against random spatial phase perturbations using ground glass scattering plates. The study indicates the possibility of using cost-effect sources such as ASE light source for generating high-dimensional spatiotemporal light fields. These fields may find applications in the fields that require resilience to both temporal noise and spatial scattering, such as reliable optical communications and advanced imaging in complex environments.

**Acknowledgments** We acknowledge financial support from the National Key Research and Development Program of China [Grant No. 2025YFF0515101 (Q.Z)], National Natural Science Foundation of China (NSFC) [Grant Nos. 12434012 (Q.Z.), 62535013 (Q.Z.) and 12474336 (Q.C.)], the Shanghai Science and Technology Committee [Grant Nos. 24JD1402600 (Q.Z.) and 24QA2705800 (Q.C.)], National Research Foundation of Korea (NRF) funded by the Korea government(MSIT) [Grant No. RS-2026-25475231 (A.C.)], Global – Learning & Academic research institution for Master's·PhD students, and Postdocs (LAMP) Program of the National Research Foundation of Korea [Grant No. RS-2023-00301938 (A.C.)], IITP (Institute of Information & Communications Technology Planning & Evaluation)-ITRC(Information Technology Research Center) grant funded by the Korea government (Ministry of Science and ICT) [IITP-2026-RS-2020-II201606 (A.C.)].